\documentclass{article}

\usepackage{hyperref}

 \usepackage{amssymb}
 \usepackage{amsmath}

\usepackage{amsthm}

\newtheorem{post}{Postulate}[section]
\newtheorem{thm}{Theorem}[section]

\newcommand{\la}{\left\langle}
\newcommand{\ra}{\right\rangle}
\newcommand{\lp}{\left(}
\newcommand{\rp}{\right)}

\title{Born's Rule From The Principle of Unitary Equivalence}

\author{Fritiof Wallentin \\ 
 Department of Mathematics, Linnaeus University, \\
 V\"axj\"o, Sweden. \\ e-mail:  fritiof.persson@lnu.se }

\begin{document}

\maketitle

\begin{abstract}
Complex phase factors are viewed not only as redundancies of the quantum formalism but instead as remnants of unitary transformations under which the probabilistic properties of  observables are invariant. It is  postulated that a quantum observable corresponds to a unitary representation of an abelian Lie group, the irreducible subrepresentations of which correspond to the observable's outcomes.  It is shown that this identification agrees with the conventional identification as self-adjoint operators. The upshot of this formalism is that one may 'second quantize' the representation to which an observable corresponds, thus obtaining the corresponding Fock space representation. This Fock space representation is then also identifiable as an observable in the same sense,  the outcomes of which are naturally interpretable as ensembles of outcomes of the corresponding non-second quantized observable. The frequency interpretation of probability is adopted, i.e. probability as the average  occurrence, from which  Born's rule is deduced by enforcing the notion 'average' to such that are invariant under the second quantized unitary representation which defines the quantum observable to which the initial state is an outcome. The enforcement of this invariance is an application the principle referred to as the {\it Principle of unitary equivalence}.
\end{abstract}


\section{Introduction}
Quantum mechanics is in this article viewed primarily as a theory of probability which emphasizes the actual experiment context. To explain this view further some heuristics of a generic probability theory will now be discussed. This is not to be viewed as rigorous mathematical definition of a generic probability theory. it will however serve as a good place for introducing concepts and notation which are crucial for this article's reformulation of quantum mechanics such that Born's rule becomes a theorem rather than a postulate.

A generic experiment  consists of some given initial condition $\psi$ with respect to which some measurement associated to an observable $A$ is performed having a range of possible outcomes $a$'s. Typically one is interested in measuring the probability distribution of the outcomes of $A$ given such an initial condition $\psi$. This is done by performing many many trials of the experiment. This results in a long sequence of measurement results of $A$,
\begin{equation} \label{ds}
(a_1,a_2,\ldots).
\end{equation}
From this the probability of an outcome $a$ is estimated by calculating its corresponding series of frequencies of occurrence, i.e.
\begin{equation} \label{f}
\lp \nu_1(\alpha),\nu_2(\alpha),\ldots\rp,
\end{equation}
where
\begin{equation}
\nu_N(\alpha)=\frac{1}{N}\sum_{n=1}^N\delta_{\alpha,\alpha_n}
\end{equation}
with $\delta$ denoting the Kronecker delta function. Such a series (\ref{f}) series is said to correspond to a certain probability if by performing a large enough number of trials of the experiment the elements of the series seems to be getting closer and closer to each other. The series is in other words a Cauchy sequence. Two series of frequencies of occurrence, irrespective from what experiment they came, are said to represent the same probability if they are equivalent as Cauchy sequences. So a probability is a label of an equivalence class of Cauchy convergent sequences, i.e an element of a completed metric space. Notice that a frequency of occurrence is the same as an average of occurrence

Just as a measurement outcome of the experiment must be something recordable, something the experimenter can unambiguously say to have occurred, so must the initial condition. This requires the initial condition to be identifiable in terms of a set of fixed observable parameters, i.e. some kind of calibration or selection. This could for instance selection for some fixed mass. But the selection could also be probability distribution. For instance, when doing some kind of poll on how people would vote if it were an election today, without having the actual election, one needs to pick a configuration of people to include in the poll that represents the whole population. One  selects an initial condition of the poll such that every relevant part of the population is represented in accordance to some desired proportions. So the initial conditions is fully specified by this (probability) distribution of people, it does not really matter which exact configuration of people are the poll then is performed upon. In fact, in order for such a poll to have any utility it is necessary that the exact configuration does not matter (too much). But note, the equivalence of configurations with respect to this poll is context specific, it is for the purpose of predicting the outcome of an actual election. This context-dependence of initial conditions will be a guiding principle in this article. In order to specify in which sense lets consider how one could quantum mechanically represent an initial conditions that is specified in terms of some probability distribution. Consider a specification of an initial conditions $\psi$ via a measured probability distribution $p(a=A|\psi)$ of outcomes $a$ of an observable $A$. $\psi$ can then be specified in terms of conventional quantum mechanics as
\begin{equation}
\left | \psi \ra=\sum_a e^{i\theta_a}|\la a|\psi\ra|\left | a \ra,
\end{equation}
since by Born's rule
\begin{equation}
p(a=A|\psi)=|\la a|\psi\ra|^2.
\end{equation}
However, this specification can only be done up to a set of complex phases $\left \{e^{i\theta_a}\right \}_a$. The ambiguity of the phases $\{e^{i\theta_a}\}_a$ can be thought of as being analogous to the equivalence of specific configurations of people in the example above of the poll. This kind of interpretation of the invariance of the complex phases will plays a crucial part in this article. And of course, since the purpose of this article is to derive Born's rule this type of equivalence will have to be motivated through some other means. In short, with out having to refer to any specific theory of probability, the up shot is that $\psi$ must be specifiable as an outcome of some observable $\Psi$.

Coming back to the generic notion of probability,  the outcomes of relevance for calculations of frequencies of occurrences are arbitrary long sequences (\ref{ds}) of outcomes of $A$. The observable associated to these outcomes is denoted $A_\infty$, which will be referred to as the {\it ensemble observable}  of a $A$. Notice that there is in this setting also an associated ensemble observable $\Psi_\infty$ to $\Psi$. One whose sequence of of outcomes is in the experiment fixed to only those of the form
\begin{equation}
(\psi,\psi,\ldots).
\end{equation}
Since the utility in the outcomes of any $A_\infty$ comes from calculating frequencies of occurrence, with such being independent of the actual order in which the outcomes occurred, outcomes of $A_\infty$'s are assumed to be invariant under permutations of  the outcomes of $A$ of which it consists. The outcomes of ensemble observables will from now on be referred to as {\it ensemble outcomes}. 

in abstract terms the counting of occurrence of an outcome $a$ of $A$ in outcomes of $A_\infty$ simply corresponds to a function $N_a$. We will refer to such functions as {\it counting functions}. Every ensemble outcome is then completely specified by the values of the functions $N_a$'s. Working formally, a calculation of probability is then something like taking the limit of the function $N_a$, 
\begin{equation} \label{1}
\frac{N_{a_0}}{\sum_{a}N_{a}}.
\end{equation}
'Taking a limit' is itself a function. Hence taking the limit of frequencies of occurrence will be seen as a function $\la\cdot\ra$ that satisfies
\begin{equation}  \label{2}
\la \sum_{a}N_{a} \ra=\sum_{a}\la N_{a} \ra=1
\end{equation}
and
\begin{equation} \label{3}
\la N_{a} \ra\geq 0.
\end{equation}
Furthermore, since the initial condition is set to  $\psi$, $\la\cdot\ra$ should be such that
\begin{equation} \label{4}
\la N_\psi\ra=\delta_{\psi,\psi'},
\end{equation}
where $\psi'$ is any other possible outcome of $\Psi$. Hence the extra label of $\psi$ is added so that the 'taking a limit' function is instead denoted $\la\cdot\ra_\psi$. The value of
\begin{equation} \label{5}
\la N_{a_0} \ra_\psi
\end{equation}
is then interpreted as the probability of outcome $a$ given initial condition $\psi$. The functions $\la\cdot\ra_\psi$ will be referred to as {\it averages}, since in the frequency interpretation of probability probabilities are calculated as the average number of outcomes. In addition of requirement (\ref{4}) $\la\cdot\ra_\psi$ should be strongly related to $\psi$ since a probability of outcome depends strongly on the initial condition. That $\la\cdot\ra_\psi$ should respect some defining feature of $\psi$, they should be {\it naturally} related. 

Now lets summarize and clarify some key points for later referral. The generic probability theory is viewed as containing:
\begin{enumerate}
\item A notion of {\it observables} such that to each observable $A$ there is a corresponding ensemble observable $A_\infty$ both being observables in the same sense.
\item Associated to each ensemble observable $A_\infty$ there are counting functions $N_a$'s, one associated to each outcome of $A$. 
\item To each initial condition $\psi$ there is some naturally related average $\la\cdot \ra_\psi$ satisfying conditions (\ref{1}) - (\ref{4}).
\end{enumerate} 

Quantum mechanics will in this article be viewed as particular representation of this just presented proposed generic structure of a probability theory. By formulating it in these terms it will in fact be shown that the usually postulated Born's rule will instead be derived, shown to equal the number given by  (\ref{5}). As already hinted at, this will be  done by providing the often ignored complex phases with a more rewarding meaning.  If they instead of being viewed as redundancies of the formalism are identified as all possible unitary transformations under a quantum observable is invariant, the notions of quantum observables can be restated as unitary representation of  abelian Lie groups, each possible outcome of the observable then corresponding to an irreducible subrepresentation. This invariance associated to quantum observables will be referred to as them satisfying the {\it principle of unitary equivalence}.  Furthermore, non-commuting self-adjoint operators have different unitary transformations under which they are invariant. This gives a way of expressing non-commutativity in the language of group representation theory. By this, this alternate view of quantum observables will have been shown to be equivalent to the conventional one  as self-adjoint operators.  This  together with further discussion on what physical interpretations one can assign to the principle of unitary equivalence is the contents of section \ref{relative phase}. 
In section \ref{observables} quantum mechanics will be (re)postulated in using this representation theoretic approach in a way that fits the above structure of a proposed generic probabilityt theory. Quantum observables will be identified as  abelian Lie groups unitarily represented  on Hilbert spaces $\mathcal{H}$. Each quantum observable will hence have an associated ensemble observable which is given by the induced representation on the symmetric Fock space $\mathcal{F}_\vee(\mathcal{H})$. This observable will be interpreted as the corresponding ensemble observable. So the ensemble observable is the second quantized\footnote{The phrase 'second quantization' is rejected by some physicist, e.q. Steven Weinberg \cite{Weinberg}. This since  quantization has already occurred in turning Poisson brackets to operator commutators. Second quantization is merely representing this obtained operator algebra on a Hilbert space. The term 'second quantization' comes from the misconception that  in quantum field theory the fields which are quantized are themselves wave-functions. They are not, they are classical fields turned into operator valued fields. The term will here exclusively refer the Fock space representation corresponding to a representation.}  version of the original observable. $\mathcal{F}_\vee(\mathcal{H})$ comes naturally  equipped with counting operators $N_a$'s. Namely the from quantum field theory well-known  number operators. By imposing the averages $\la\cdot\ra_\psi$ to be invariant under the action of second quantized version of $\Psi$ it is then shown that Born's rule follows. This imposing of invariance will be seen as a particular manifestation of the principle of unitary equivalence in terms of probability. Section \ref{discussion} is a summary of the basic key points. This article ends with section \ref{gauge origin?}, an appendix in which it is speculated if the principle of unitary equivalence has its origin in gauge theory.

Before ending this introduction it is well worth emphasizing that the author views Kolmogorov probability \cite{Khrennikov1} as also fitting this notion of a generic probability theory. One for which Hilbert spaces are exchanged for measure spaces, observables are random variables, ensemble observables are independent and  identically distributed ({\it i.i.d}) random variables  and the sums of i.i.d random variables are analogous to the counting functions. Within this view the Law of large numbers is a theorem that shows that how to calculate probabilities from the measure of the measure space.\footnote{This view is different than the conventional one in Kolmogorov probability. But according to the author limits of frequencies of occurrence is \underline{the} interpretation of probability that is ultimately always applied in the empirical sciences. Thus with out the Law of large numbers there is no theoretical justification for identifying the normalized measure as \underline{the} probability. The normalized measure obtains its interpretation as measure of probability through the Law of large numbers because the latter is a way of calculating limits of frequencies of occurrence.  Put shortly, the law of large numbers is the justification for referring to a normalized measure as a 'probability measure'. The author claims that with out the law of large numbers the identification of the normalized measure as the probability would be just as mysterious as Born's rule. See \cite{K1} for more discussions regarding interpretations of the Law of large numbers.} Similarly Born's rule will here be shown to be a theorem showing how to calculate probabilities by using the inner product of the Hilbert space.

\section{Complex Phases in Quantum Mechanics} \label{relative phase}
This section will pave the way for the formulation of quantum mechanics which will be applied in section \ref{observables} when deducing Born's rule. This will be done by highlighting the importance of complex phases in Quantum mechanics, how they are not redundant but really crucial.

For each $j=1,2$ let $A^j$ be a non-degenerate  self-adjoint operator on a complex $N$-dimensional Hilbert space $\mathcal{H}$. Let $\{\left  |a^j_n\ra\}_{n=1}^N$ denote the eigenbasis of $A^j$ and $\{a^j_n\}_{n=1}^N$ the corresponding eigenvalues.  Consider any initial state $\left |\psi \ra\in \mathcal{H}$. Then, 
\begin{equation} \label{a}
\left | \psi\ra = \sum_{n=1}^N c_{a_n}e^{i\theta_{a^j_n}}\left | a^j_n\ra  ,
\end{equation}
with  $\{c_{a^j_n}\}_{n=1}^N\subset \mathbb{R}$. As is well known, multiplying  $\left | \psi\ra$ by an arbitrary complex phase makes no observably detectable difference upon measurement of $A^j$.  But that is not the only symmetry present.  According to Born's rule the probabilities of outcomes of $A^
j$ are invariant under any change of the $\theta_{a^j_n}$'s in (\ref{a}).  Such symmetries  are here referred to as {\it relative phase symmetries}. 

Changes of relative phases  can be formalized as actions of unitary operators
\begin{equation} \label{ua}
U_{A^j}(\overline{\theta})=\sum_{n=1}^Ne^{\theta_n}\left | a^j_n\ra\la a^j_n\right|,
\end{equation}
where 
\begin{equation} \label{th}
\overline{\theta}=(\theta_1,\ldots, \theta_N)\in  \mathbb{R}^N.
\end{equation} 
If $A^1$ and $A^2$ are non-commuting, the actions (\ref{ua}) of $A^1$ on $\left |\psi\ra$ do not leave the probabilities of outcomes of a $A^2$ invariant, and vice versa. Such effects are in fact what shows up as interference in double-slit type experiments. Through  a straightforward calculation,
\begin{equation} 
U_{A^j}(\overline{\theta}_1+\overline{\theta}_1)=U_{A^j}(\overline{\theta}_1)U_{A^j}(\overline{\theta}_2).
\end{equation}
Hence the map $U_{A^j}$ induces  a representation of $\mathrm{U}(1)^N$ on $\mathcal{H}$.  Let $\mathcal{H}_{A^j}$ denote this representation.
The representations $\mathcal{H}_{A^1}$ and $\mathcal{H}_{A^2}$  are unitarily equivalent since for all choices of  $\theta_{a^1_{n}a^2_{\sigma(n)}}$'s in $ \mathbb{R}$ and any choice of permutation of $N$ elements, $\sigma$, the operator
\begin{equation} \label{T}
 T_\sigma =\sum_{n=1}^N e^{i\theta_{a^1_{n}a^2_{\sigma(n)}}}\left | \left. \left. a^2_{\sigma(n)}\ra \right \langle a^1_n\right |
\end{equation}
 is an intertwining\footnote{Recall: An operator $O:\mathcal{H}_{A^1}\rightarrow \mathcal{H}_{A^2}$ is intertwining if
$$ OU_{A^1}(\overline{\theta}) =U_{A^2}(\overline{\theta})O$$
for all $\theta_1,\theta_2\in\mathbb{R}$.} 
isomorphism. In fact all intertwining isomorphism between $\mathcal{H}_{A^1}$ and $\mathcal{H}_{A^2}$ are of that form. But, since for any $\sigma$,
\begin{equation}
|\la a^2_m\right |T_\sigma \left |a^1_n \ra|^2 = \delta_{m,\sigma(n)} \neq |\la a^2_m|a^1_n
 \ra|^2,
\end{equation}
these intertwining operators do not preserve probabilities of outcomes, the latter being of great importance in quantum mechanics. The only unitary isomorphisms preserving probabilities of outcomes of $A^1$ and $A^2$ are those which are the identity map times some phase factor. So  $\mathcal{H}_{A^1}$ and $\mathcal{H}_{A^2}$ are quantum mechanically equivalent if the identity map is intertwining.  Notice that {\it quantum mechanically equivalence} is just another phrase for commutativity. However as will be come clearer later the first term is more appropriate term in this article since quantum mechanics will not be explicitly stated in terms of observables being self-adjoint operators but in terms of unitary representation. With that in mind the upshot is that there is a natural group representation theoretic way of expressing non-commutativity. Furthermore, that quantum observables indeed can be expressed in terms of representation theory follows from  $\mathcal{H}_{A^j}$ being decomposable into irreducible subrepresentations of $U_{A^j}$ as
\begin{equation} \label{RF}
\mathcal{H}_{A^j}= \bigoplus_{n=1}^N\mathrm{span}_{\mathbb{C}}\left\{\left | a^j_n\ra\right\}.
\end{equation}
 That is, the decomposition (\ref{RF}) is identical the decomposition with respect to the eigenstates of the corresponding  self-adjoint operator. So by attaching suitable real numbers, i.e the corresponding eigenvalues, to each irreducible subrepresentation of $\mathcal{H}_{A^j}$ the self-adjoint operator $A^J$ is reconstructed.  However, this pairing seems ambiguous since there seems to be no restriction on what 'eigenvalues' that are viable choices. This issue will be discussed next.
 
From now on the  extra $j$-index will be dropped, instead simply denoting the  generic quantum observable  as $A$ and its  possible outcomes as $a_n$. By Schur's lemma \cite{Woit} any unitary (faithful) representation $U$ of $\mathrm{U}(1)^N$ can be represented as 
 \begin{equation} 
U(\overline{\theta})=\sum_{n=1}^N e^{i k_n\theta_n} p_n
\end{equation}
with $\{k_n\}_{n=1}^N\subset \mathbb{Z}$ and  $\{p_n\}_{n=1}^N$ being a complete set of orthogonal projections on the considered Hilbert space. Then for any path in $\mathrm{U}(1)^N$ parameterized by  a smooth function $\overline{\theta}(t)$ for which $\overline{\theta}(0)$, there is a unique associated self-adjoint operator $X_{\overline{\theta}}$ defined as
\begin{equation} \label{ob1}
-i\left.\frac{d}{dt}\right|_{t=0}U(\overline{\theta}(t))=\sum_{n=1}^N k_n\dot{\theta}_n(0) p_n.
\end{equation}
In particular, if  the path is such that
\begin{equation} \label{ob2}
k_n\dot{\theta}_n(0)=a_n
\end{equation}
for any $n=1,\ldots ,N$, then $X_{\overline{\theta}}$ has the same eigenvalues as $A$. So every 'direction' in $\mathrm{U}(1)^N$, i.e. Lie algebra element, corresponds to a quantum observable in terms of a self-adjoint operator. Moreover, since (\ref{ob2}) is solvable for any real valued $a_n$, every choice of values of outcomes is valid. All these quantum observables are furthermore quantum mechanically equivalent. So not so surprisingly merely stating that a  observable is unitary representation of an abelian Lie group gives no information of the actual values of observable outcomes. But this is no different that in the conventional case where an observable is merely a self-adjoint operator or in the classical Kolmogorov case where an observable merely is a random variable. The restrictions must come from elsewhere.

The reduction of possible choices of self-adjoint operators to which a unitary representation corresponds can for instance be done by introducing another quantum observable $B$ for which there is some previously  known relation to $A$. Such a relation could be identified through:
\begin{enumerate}
\item statistics. That is, by performed measurements one can obtain the probabilities 
\begin{equation}
P(a_m|b_n).
\end{equation}
By applying Born's rule one has then obtained the numbers
\begin{equation}
|\la a_m|b_n\ra|^2=P(a_m|b_n).
\end{equation}
Then one knows the relation between the different eigenbases up to a set of phase factors $\{e^{i\phi_{a_m,b_n}}\}_{m,n}$, i.e
\begin{equation} \label{co}
\begin{split}
\left | a_m\ra &=\sum_{n=1}^Ne^{\theta_{a_m,b_n}}\sqrt{P(a_m|b_n)} \left | b_n\ra ,\\ 
\left | b_m\ra &=\sum_{n=1}^Ne^{-i\theta_{a_n,b_m}}\sqrt{P(a_n|b_m)}  \left | a_n\ra .
\end{split}
\end{equation}
In other words, fixing the eigenbasis of $B$ to some preferred one, then for any with respect to (\ref{co})  consistent choice of representation $U_A$, the representation 
\begin{equation} \label{d}
\overline{\theta}\mapsto U_B(\overline{\phi})^*U_A(\overline{\theta})U_B(\overline{\phi}),
\end{equation}
for any fixed $\overline{\phi}$, is just as consistent. So having fixed the representation of $B$ the representation of $A$ is only uniquely defined up to unitary transformations $U_B(\overline{\phi})$. But the decomposition of $A$ into irreducible subrepresentations remains identical under any transformation (\ref{d}).
\item something like quantization of a classical theory, where from some other mathematical structure a relation between the observables is known. However,  even when  quantizing a classical theory there is still room for equivalence under unitary transformations. Consider for instance  the canonical commutation relations
\begin{equation} \label{com}
\left[P,Q \right]=i\hbar I.
\end{equation}
Even when this only involves a finite number of degrees of freedom Stone-von Neumann's theorem \cite{Hall} only provides a unique representation up to a class of unitary transformations. In particular, since for any $s\in \mathbb{R}$,\footnote{Which can be shown by considering the series expansion of $e^{-iQs}$  and utilizing (\ref{com}) term-wise.}
\begin{equation} \label{t}
e^{iQs}Pe^{-iQs}=P+s, 
\end{equation}
it follows that
\begin{equation}
\left[e^{iQs}Pe^{-iQs},Q \right]=\left[P,Q \right].
\end{equation}
 This essentially means that there is an arbitrariness to where 'zero momentum' is. So this equivalence under unitary transformations can be seen as similar to that of not being able to fix a unique self-adjoint operator to a unitary representation $\mathcal{H}_A$. Furthermore, (\ref{t})  is not in essence a quantum mechanical phenomena. It is a integral part in classical mechanics as well, due to Galilean covariance\footnote{\cite{Schwichtenberg} is a good reference for a distinction between {\it invariance} and {\it covariance}. In simple terms, {\it invariance} means that the quantity of a certain quality remains constant under the considered transformation while  {\it covariance} means that the rule for calculating quantities of this certain quality remains constant. Note that, in this sense {\it covariance} can be stated as an invariance, but as an invariance on the space of 'calculation rules'. This is a reason why these concepts can be hard to distinguish sometimes. }   In basic terms, in classical physics covariance under certain transformations  means  hat the \underline{laws} of physics are \underline{invariant} under these transformation.  So reasoning by analogy it seems reasonable to have a corresponding statement in quantum mechanics with regards to unitary equivalences. 
\end{enumerate}

To emphasize, the purpose of introducing points 1. and 2. above was to exemplify that for  standard ways of constructing a quantum mechanical representation of some set of observables the unitary equivalences are nonetheless  present. To put the reasoning at the end of point 2. in  different terms, for this unitary equivalence of representations to not cause ambiguities, any way of obtaining observables quantities should be invariant under this same unitary equivalence. Lets formally refer to this as the {\it principle of unitary equivalence}, {\it PUE} for short.  The postulation of probability in quantum mechanics satisfying PUE  will in this article be a key ingredient in the derivation of Born's rule.\footnote{See Postulate \ref{invprob}.}

Lastly, the derivation of Born's rule in this article can also be seen as a proof of consistency. In a similar sense that the Law of large numbers is in Kolmogorov probability. But as stated in the introduction, the author does not agree with that interpretation of the Law of large numbers. Since it is through  limit of frequencies of occurrence that probabilities are empirically identified. The author views proofs of Law of large number-type theorems as  providing 'simpler' rules for calculating probabilities within the theory.

\section{The Quantum Mechanical Representation of Experiments} \label{observables}

In this section the postulation of quantum mechanics will be done such that it matches the blue print of a heuristic generic probability theory as presented in the introduction. The identification of observables will be as unitary representations of abelian Lie groups which in section \ref{relative phase}   was shown to  be equivalent to the conventional postulation \cite{K1} as self-adjoint operators. The difference with standard postulation is that Born's rule will here not be postulated. Applying the terminology introduced in the introduction, it will instead be postulated that the rule $\la\cdot\ra_\psi$ for taking limits of frequencies of occurrence satisfies PUE.

Motivated by the discussion in section \ref{relative phase} the following is postulations are made:

\begin{post}
An observable $A$ with possible outcomes $\{a_n\}_{n=1}^N$ is  quantum mechanically represented as a unitary representation  $U_A$ of an abelian Lie group $G_A$ such that each outcome $a_n$ can be associated to one and only one irreducible subrepresentation of $U_A$. 
\end{post}

Notice that each outcome $a_n$ has a unique orthogonal one-dimensional  projection $p_{a_n}$ associated to it. Namely the one that projects onto the irreducible subrepresentation to which $a_n$ corresponds.

\begin{post}
A  quantum mechanical representation of an experiment $\mathcal{E}_{\Psi\rightarrow A}$ consists of a Hilbert space $\mathcal{H}$ on which $A$ and $\Psi$  are both quantum mechanically represented. Generically denotes as $(\mathcal{H},U_A,U_{\psi})$.
\end{post}

The task is now to identify the corresponding quantum representation of the ensemble observables $A_\infty$ and $\Psi_\infty$ making up the ensemble experiment $(\mathcal{H}_\infty,U_{A_\infty},U_{\Psi_\infty})$. This will be done by identifying the ensemble experiment as  the 'second quantized' version of $(\mathcal{H},U_A,U_{\psi})$. But before that is postulated lets first motivate why and show that this indeed satisfies the heuristics of a generic probability theory as presented  in the introduction.

Let $\mathcal{H}_\infty$ be the symmetric Fock space\footnote{For mathematical details about  Fock spaces and second quantization  the reader is referred to \cite{Attal} and  \cite{JohnnyT}.}, 
\begin{equation}
\mathcal{F}_\vee\lp \mathcal{H}\rp=\bigoplus_{N=0}^\infty \bigvee_{n=0}^N \mathcal{H},
\end{equation}
$U_{A_\infty}=\Gamma(U_A)$ and $U_{\Psi_\infty}=\Gamma(U_\Psi)$, where $\Gamma$ is  defined such that for any operator $O$ on $\mathcal{H}$,
\begin{equation}\label{afs}
\Gamma(O) \vee_{n=1}^N \left | \phi_n \ra=\vee_{n=1}^N  O  \left | \phi_n \ra,
\end{equation}
and linearly extended. Let $g_A(t)$ be a path in $G_A$ such that $g_A(0)=I$ and for any $n$ let $\left | a_n \ra$ denote a generic normalized representative of the irreducible subrepresentation in $\mathcal{H}$ to which the outcome $a_n$ corresponds.\footnote{This is because it can be used to $\left | a_n \ra$ represent the orthogonal projection operator $p_{a}$ as
$\left | a_n\ra \la a_n\right |$.  So no motivation based on probability is needed.} Then since $\left | a_n \ra$ is an element of a unitary irreducible subrepresentation, there is a $(\mathbb{R},+)$-group homomorphism $\theta_n$ such that
\begin{equation}
U_{A}(g_A(t))  \left | a_n \ra = e^{i\theta_n (t)}  \left | a_n \ra.
\end{equation}
Hence 
\begin{equation}
\begin{split}
-i\left. \frac{d}{dt}\right|_{t=0} U_{A_\infty}(g_A(t)) \vee_{n=1}^N \left | a_{k_n} \ra & =  -i\left. \frac{d}{dt}\right|_{t=0}\vee_{n=1}^N   U_{A}(g_A(t))  \left | a_{k_n} \ra \\
& =  \lp \sum_{n=1}^N \dot{\theta}_{k_n}(0) \rp \vee_{n=1}^N    \left | a_{k_n} \ra ,
\end{split}
\end{equation}
where the second equality follows from applying Leibniz rule.
So each element $\vee_{n=1}^N \left | a_{k_n} \ra$ spans an irreducible subrepresentation of $U_{A_\infty}$. Furthermore, these elements have a natural identification as sequences of outcomes of measurements of $A$ which are symmetric under permutations of the order of the outcomes. Which, as stated in the introduction, is what is required of outcomes of $A_\infty$.  Hence the following postulate is stated:

\begin{post}
If $A$ is  as a quantum observable  the unitary representation $(\mathcal{H},U_A)$, then its ensemble observable $A_\infty$ is quantum mechanically represented by the second quantization of $(\mathcal{H},U_A)$, i.e. as
\begin{equation}
\lp \mathcal{F}_\vee\lp\mathcal{H}\rp,\Gamma\lp U_A \rp \rp.
\end{equation}
\end{post}

For the purpose of  identifying the counting functions $N_a$'s first notice that
\begin{equation}
\lp \sum_{m=1}^M \dot{\theta}_{k_m}(0) \rp \vee_{m=1}^M    \left | a_{k_m} \ra  
  =  \sum_{n=1}^N \lp \sum_{m=1}^M \delta_{n,k_m}\dot{\theta}_{k_m}(0) \rp \vee_{m=1}^M    \left | a_{k_m} \ra.
\end{equation}
So by making the totally valid choice  of the path $g_A(t)$ such that each $\dot{\theta}_{n}(0)=\delta_{n_0,n}$, one obtains
\begin{equation}
 \sum_{n=1}^N \lp \sum_{m=1}^M \delta_{n,k_m}\dot{\theta}_{k_m}(0)\rp =   \sum_{m=1}^M \delta_{n_0,k_m},
\end{equation}
i.e for any given number of trials $M$ it counts the total number of occurrences of $a_{n_0}$. So $-i\left. \frac{d}{dt}\right|_{t=0} U_{A_\infty}(g_A(t)) $ with such a choice of path $g_A(t)$ corresponds to the number operator $N_{a_{n_0}}$ as  known from quantum field theory \cite{Peskin}. Of course an analogous analysis holds for $U_{\Psi_\infty}$ as well. Furthermore, by
defining the operation $d\Gamma$ as
\begin{equation}
d\Gamma(O) \vee_{m\in I} \left | \phi_m \ra=\sum_{n\in I}\vee_{m\in I\backslash \{n\}}\left | \phi_{m} \ra\vee O \left | \phi_{n} \ra,
\end{equation}
for any operator $O$ on $\mathcal{H}$,  it follows that
\begin{equation}
d\Gamma\lp p_{a_n}\rp =N_{a_{n_0}}.
\end{equation}
In fact, for any one-parameter group $U(t)$ on $\mathcal{H}$,
\begin{equation} \label{lin}
\left. \frac{d}{dt}\right|_{t=0} \Gamma\lp U(t)\rp =d\Gamma \lp \left. \frac{d}{dt}\right|_{t=0} U(t)\rp.
\end{equation}
 Moreover, for any operators  $O_1$, $O_2$ and $O_3$ on $\mathcal{H}$,
\begin{equation}
\Gamma\lp O_1\rp\Gamma\lp O_2\rp\Gamma\lp O_2\rp=\Gamma\lp O_1O_2O_3\rp,
\end{equation}
which follows from the definition of $\Gamma$ in (\ref{afs}). From this, (\ref{lin}) and since 
\begin{equation}
 -i \left. \frac{d}{dt}\right|_{t=0} U_A(g_A(t)) = p_{a_{n_0}} ,
\end{equation}
 it follows that
\begin{equation}
U^*_{\Psi_\infty}(g_\Psi)N_{a_n}U_{\Psi_\infty}(g_\Psi)  =d\Gamma \lp U^*_{\Psi}(g_\Psi) p_{a_{n_0}} U_{\Psi}(g_\Psi)\rp .
\end{equation}
In addition 
\begin{equation}
\begin{split}
U^*_{\Psi}(g_\Psi) p_{a_{n_0}} U_{\Psi}(g_\Psi) &=\sum_{k=1}^N |\la a_{n_0}| \psi_k \ra|^2 p_{\psi_k} \\
&+ \sum_{k\neq m, k,m=1}^Ne^{i(\theta_k(g_\Psi)-\theta_m(g_\Psi)}p_{\psi_m}p_{a_{n_0}}p_{\psi_k}.
\end{split}
\end{equation}
So
\begin{equation} \label{b}
\begin{split}
U^*_{\Psi_\infty}(g_\Psi)N_{a_{n_0}}U_{\Psi_\infty}(g_\Psi) &=\sum_{k=1}^N |\la a_{n_0}| \psi_k \ra|^2 N_{\psi_k} \\
& + \sum_{k\neq m, k,m=1}^N e^{i(\theta_k(g_\Psi)-\theta_m(g_\Psi)} d\Gamma\lp p_{\psi_m}p_{a_{n_0}}p_{\psi_k}\rp.
\end{split}
\end{equation}
Now suppose there exists a linear function $\la\cdot\ra_{\psi_{n_0}}$ on the space of (densely defined) operators on $\mathcal{H}_\infty$, which in addition is invariant under the action of $U_{\Psi_\infty}$ and such that
\begin{equation}
\la N_{\psi_{k}}\ra_{\psi_{k_0}}= \delta_{k,k_0}.
\end{equation}
The invariance under $U_{\Psi_\infty}$ means that
\begin{equation} \label{da}
\la N_{a_n}\ra_{\psi_{k_0}}=\la U^*_{\Psi_\infty}(g_\Psi)N_{a_n}U_{\Psi_\infty}(g_\Psi)\ra_{\psi_{k_0}},
\end{equation}
and hence 
\begin{equation} \label{refa}
\la d\Gamma\lp p_{\psi_m}p_{a_{n_0}}p_{\psi_k}\rp \ra_{\psi_{k_0}}=0,
\end{equation}
since by (\ref{da}) no dependence on the phase factors $e^{i(\theta_k(g_\Psi)-\theta_m(g_\Psi)}$ from (\ref{b}) can be present.  So by (\ref{b}-\ref{refa})  one ends up with 
\begin{equation} \label{born}
\la N_{a_n}\ra_{\psi_{k_0}}= |\la a_{n_0}| \psi_k \ra|^2,
\end{equation}
i.e. Born's rule.
Furthermore, such averages $\la\cdot\ra_{\psi_{n_0}}$ do exist. For instance, any
\begin{equation}
\frac{\la \vee_{n\in I}\left | \psi_{i_n}\ra | \cdot|  \vee_{n\in I}\left | \psi_{i_n}\ra \ra}{
\la \vee_{n\in I}\left | \psi_{i_n}\ra | N_{\psi_{k_0}} |\vee_{n\in I}\left | \psi_{i_n}\ra \ra}
\end{equation}
with  $i_n=k_0$ for all $n$,  will do. 

Hence the following postulates are stated:
\begin{post}
The quantum mechanical representation of a counting function $N_a$ associated to a quantum representation of the observable $A$ is given by
\begin{equation}
N_a=d\Gamma(p_a).
\end{equation}
\end{post}

\begin{post} \label{invprob}
The quantum mechanical representation of $\la\cdot\ra_{\psi}$ associated to an outcome of the quantum mechanical representation $(\mathcal{H},U_\Psi)$ of the observable $\Psi$ is a $\Gamma(U_\Psi)$-invariant linear functional on the space of densely defined operators on $F_\vee\lp\mathcal{H}\rp$, i.e such that
\begin{equation}
\la \Gamma(U_\Psi)^* \lp\cdot\rp \Gamma(U_\Psi)\ra_\psi=\la\cdot\ra_{\psi}.
\end{equation}
In other words, $\la\cdot\ra_{\psi}$ satisfies PUE under $\Gamma(U_\Psi)$.
\end{post}

Notice that Postulate \ref{invprob} also specifies  the notion of {\t natural relationship} between $\psi$ and $\la\cdot\ra_{\psi}$ that was stated as a requirement in the generic probability theory from the introduction. That is, they satisfy the 'same' symmetry.

As has been shown through the calculations leading to (\ref{born}), Born's rule is a theorem in this postulation of quantum mechanics.
\begin{thm} \label{bornthm}
In the quantum mechanical representation of an experiment 
$$\mathcal{E}_{\psi\in \Psi\rightarrow A}$$
 the probability $P_{\psi}(a_n)$ of obtaining $a_n$ is given by Born's rule, i.e.
\begin{equation}
P_{\psi}(a_n)=|\la a_n|\psi\ra|^2.
\end{equation} 
\end{thm}

\section{Summary} \label{discussion}

\begin{enumerate}
\item A Quantum mechanical observable $A$ a is unitary representation
\begin{equation}
U_A:G_A\rightarrow U(\mathcal{H})
\end{equation}
 of an abelian Lie group $G_A$ on a Hilbert spaces $\mathcal{H}$, the irreducible subrepresentations of which constituting the possible outcomes of the observable. Consequently each possible outcome $a$ can be uniquely identified with the orthogonal projection operator $p_a$ to which it corresponds.
\item An experiment
\begin{equation}
\mathcal{E}_{\psi\in \Psi\rightarrow A}
\end{equation}
 for which there is an observable $\Psi$ associated to the initial condition $\psi$ and an observable $A$ associated to the measurement performed given initial condition $\psi$ consists of
 \begin{equation}
U_\Psi:G_\Psi\rightarrow U(\mathcal{H})
\end{equation}
and
\begin{equation}
U_A:G_A\rightarrow U(\mathcal{H}).
\end{equation}
\item Measurements of probabilities of outcomes in $\mathcal{E}_{\psi\in \Psi\rightarrow A}$ are done by performing many many trial of the experiment and calculating frequencies of occurrence, which for large enough number of trials are considered as probabilities.  These kinds of measurements correspond to the experiment 
\begin{equation}
\mathcal{E}^\infty_{\psi\in \Psi\rightarrow A}
\end{equation}
consisting of
  \begin{equation}
\Gamma\lp U_\Psi\rp:G_\Psi\rightarrow U\lp F_\vee\lp\mathcal{H}\rp\rp
\end{equation}
and
\begin{equation}
\Gamma\lp U_A\rp :G_A\rightarrow U\lp F_\vee\lp\mathcal{H}\rp\rp,
\end{equation}
i.e. the second quantized version of $\mathcal{E}_{\psi\in \Psi\rightarrow A}$.
\item The operator 
\begin{equation}
N_a=d\Gamma(p_a)
\end{equation}
unambiguously assigns the number of occurrence of the outcome $a$ in an outcome of $\mathcal{E}^\infty_{\psi\in \Psi\rightarrow A}$. 
\item Any function $\la\cdot\ra_\psi$  representing the calculations of probabilities in the experiment $\mathcal{E}^\infty_{\psi\in \Psi\rightarrow A}$, i.e. 
\begin{equation}
P(a|\psi)=\la N_a\ra_\psi,
\end{equation} 
must by elementary postulates of probability theory $\la\cdot\ra_\psi$ satisfy: 
\begin{equation}
\begin{split}
\la\sum_a N_a\ra_\psi&=\sum_a \la N_a\ra_\psi=1, \\
\la N_a\ra_\psi&\geq 0, \\
\la N_{\psi'}\ra_\psi &=\delta_{\psi,\psi'}.
\end{split}
\end{equation}
\item  Let any such $\la\cdot\ra_\psi$  in addition be a $\Gamma(U_\Psi)$-invariant linear functional  on the space of densely defined operators on $F_\vee\lp\mathcal{H}\rp$. It then follows Theorem \ref{bornthm}  that $\la N_a\ra_\psi$ satisfies Born's rule, i.e
\begin{equation}
\la N_a\ra_\psi=\mathrm{Tr}\left\{ p_a p_\psi\right \}.
\end{equation}
\item The $\Gamma(U_\Psi)$-invariance of $\la\cdot\ra_\psi$ is motivated by the  quantum representation of observables in general is unique only up to a class of unitary transformations, so in order for probability to remain unambiguous it should remain invariant under these. This is considered as an application of the  Principle of unitary equivalence for probability. 
\end{enumerate}

\section[Gauge theory-like Origin of the Principle of Unitary Equivalence?]{Appendix: Gauge theory-like Origin of \\ the Principle of Unitary Equivalence?} \label{gauge origin?}
In this section similarities between PUE in quantum mechanics and local gauge transformations in gauge field theory will be discussed. Just the bare minimum of gauge field theory to get the point across will be presented. For a more thorough introduction the reader is  refer to \cite{Peskin}, \cite{Folland} and \cite{Hamilton,Hamilton2,Valery}, where the first is a standard textbook on quantum field theory, the second an introduction to the subject aimed at mathematicians and the rest dealing only with classical gauge theory. 

Let $\mathcal{M}$ be a space time-manifold  and $G$ some Lie group. Consider the spaces
\begin{equation}
C^\infty(\mathcal{M}, \mathbb{C}^n)
\end{equation}
and 
\begin{equation}
C^\infty(\mathcal{M}, G).
\end{equation}
Suppose there is a  unitary representation of $G$ onto $\mathbb{C}^n$ so that one can define an action $\cdot $ of $C^\infty(\mathcal{M}, G)$ onto $C^\infty(\mathcal{M},\mathbb{C}^n)$ as 
\begin{equation} \label{gauge rep}
(g(x),\Psi(x))\in C^\infty(\mathcal{M} ,G)\times C^\infty(\mathcal{M} , \mathbb{C}^n)\mapsto g(x)\cdot\Psi(x)\in C^\infty(\mathcal{M}, \mathbb{C}^n).
\end{equation}
$C^\infty(\mathcal{M}, G)$ is referred to as the gauge group of $G$.
The elements of $C^\infty(\mathcal{M},\mathbb{C}^n)$ are not to be mistaken as quantum mechanical wave-functions. As mentioned in \cite{Woit}, the action of $C^\infty(\mathcal{M} ,G)$ is not upon a finite dimensional phase space of coordinates and momenta, so it has no interpretation in terms of ordinary quantum mechanical quantization. In gauge theory $C^\infty(\mathcal{M},\mathbb{C}^n)$ is identified as the phase space which is turned into a quantum field theory by identifying the canonical coordinates and canonically quantizing their Poisson bracket \cite{Peskin}. As such the quantized fields are more appropriately analogous position and momentum operators than wave functions.

 The dynamics of  the \textit{matter fields} $\Phi\in C^\infty(\mathcal{M},\mathbb{C}^n)$ are described by classical field equations derived as Euler-Lagrange equations from a Lagrangian
\begin{equation}
\mathcal{L}[\Phi].
\end{equation}
As explained in \cite{Hamilton}, by requiring $\mathcal{L}$ to be invariant under gauge transformations, i.e. 
\begin{equation}
\mathcal{L}[\Phi]=\mathcal{L}[g\Phi]
\end{equation}
for $g\in C^\infty(\mathcal{M}, G)$, one ensures that if a $\Phi$ is a solution to the field equations, then so is $g\Phi$. The gauge invariance of the Lagrangian is stronger than just requiring gauge invariance of the field equations. The reason physicists use this stronger requirements is that it gives a gauge invariant action functional\footnote{Solutions of the field equations correspond to critical points of the action functional.} 
\begin{equation}
\mathcal{S}[\Phi]=\int_\mathcal{M} \mathcal{L}[\Phi] dx
\end{equation}
which is a requirement in path-integral quantization. Details of such deeper structure will not be of main interest here. The focus will instead be on the interactions coming from requiring this invariance. This requirement  leads to the introduction of \textit{gauge fields} $B=(B_\mu)$,
\begin{equation}
B_\mu\in C^\infty(\mathcal{M}, \mathfrak{g}),
\end{equation}
where $\mu$ is the space-time index,
acting on the matter fields via the corresponding Lie algebra action of $ C^\infty(\mathcal{M}, G)$. These added gauge fields induces coupled non-linear field equations with the matter and gauge fields $\Phi$ and $B$. This means that there exists interactions between these. It is because of this the gauge fields are referred to as force carriers. All fundamental forces of nature are derivable as gauge theories.\footnote{Although not all can be unified in a corresponding quantum theoretic framework, e.g. grand unification and quantum gravity.}

 An elementary particle's property with respect to a gauge symmetry is given by the irreducible representation to which it corresponds. For instance, solely in terms of the strong nuclear force quarks are identified is as the fundamental representation of $\mathrm{SU}(3)$ \cite{Peskin}. With this in mind think of associating an observable $O$ with a representation of the gauge group $C^\infty(\mathcal{M}, G_O)$ on  $C^\infty(\mathcal{M}, \mathbb{C}^n)$. Given a field $\Psi\in C^\infty(\mathcal{M}, \mathbb{C}^n)$, the possible observable outcomes of $O$ correspond respectively to the irreducible components of $(\mathbb{C}^n,G_O)$. Hence a similar structure of observables as that suggested in this article has been obtained. So here  the method of section \ref{observables} could be applied to obtain the probabilistic framework of quantum theory. Note that this would give the conventional formalism of quantum mechanics on $\mathbb{C}^n$. The fields $\Phi$ would hence be 'wavefunction-valued' functions\footnote{Perhaps more correctly distributions, even.} on space-time not wavefunctions themselves. The point is that gauge theory could provide a reason for the principle of unitary equivalence in quantum mechanics. This hypothetical reason could be something like this:
\begin{itemize}
\item A measurement is an interaction. The interaction coming from enforcing some gauge symmetry of some Lagrangian $\mathcal{L}$.
\item Thus a measurement fundamentally\footnote{Since the property that is measures is defined as this interaction that is itself the measurement.} makes no distinction between different 'gauges' of its possible outcomes.
\item Hence any quantity that is obtainable through a measurement must respect this gauge symmetry.
\end{itemize}
The last point is then what corresponds to the principle of unitary equivalence, which in this article is in particular applied to calculating probabilities as limits of frequencies of occurrence. To emphasize, in this setting only classical field theories have been considered. The 'quantum weirdness' of quantum mechanics is a consequence of Born's rule. By the view of this article the 'quantum weirdness' would appear through second quantization. In this sense 'first quantization' of a gauge theory, i.e. replacing $\mathbb{C}^n$-valued field with operator valued, could in this sense still be considered 'classical'. Perhaps it could even  be made analogous to what the Koopman-von Neumann formulation \cite{Frank} is to classical mechanics? Since second quantization naturally has an ensemble interpretation associated with it, as this article argues, could such a view remove the 'quantum weirdness' from quantum mechanics?

In closure, this appendix is merely a speculation. In no way has a proper gauge theory to model quantum measurement been constructed. This sections main purpose has been to show that the notion of symmetries and irreducible representations have been successfully applied in construction of one of the best accomplishments of science, i.e. the standard model of particle physics. There, just as here, is the true nature of gauge symmetries not known. Are they just mathematical redundancies or is there deeper physics beneath? In this appendix  it has been speculated on how gauge theories may prove to be theories of measurements  for which quantum probability apply. If this could be rigorously shown, then it would seem that quantum mechanical effects stemming from Born's rule would be phenomena of the collective of all part: measured system, measuring device and the ensemble of trials of measurements; as opposed to just the measured system. This would then have a huge impact on the interpretation of quantum mechanics.

\end{document}